\documentclass[submission, Proceedings]{SciPost}

\hypersetup{
    colorlinks,
    linkcolor={red!50!black},
    citecolor={blue!50!black},
    urlcolor={blue!80!black}
}

\usepackage[bitstream-charter]{mathdesign}
\usepackage{booktabs}
\DeclareSymbolFont{usualmathcal}{OMS}{cmsy}{m}{n}
\DeclareSymbolFontAlphabet{\mathcal}{usualmathcal}
\allowdisplaybreaks

\newcommand{\seq}{\begin{subequations}}
\newcommand{\sen}{\end{subequations}}
\newcommand{\be}{\begin{eqnarray}}
\newcommand{\ee}{\end{eqnarray}}
\newcommand{\eq}{\begin{eqnarray}}
\newcommand{\en}{\end{eqnarray}}

\newcommand{\bfk}{{\bf k}_{\perp}}

\begin{document}

\begin{center}{\Large \textbf{
Proton gravitational form factors in a light-front quark-diquark model \\
}}\end{center}

% TODO: write the author list here. Use initials + surname format.
% Separate subsequent authors by a comma, omit comma at the end of the list.
% Mark the corresponding author with a superscript *.
\begin{center}
Dipankar Chakrabarti\textsuperscript{1},
Chandan Mondal\textsuperscript{2,3,4},
Asmita Mukherjee \textsuperscript{5},
Sreeraj Nair \textsuperscript{2,3,4$\star$} and
Xingbo Zhao \textsuperscript{2,3,4}
\end{center}

\begin{center}
{\bf 1} Department of Physics,
Indian Institute of Technology Kanpur, Kanpur 208016, India
\\
{\bf 2} Institute of Modern Physics, Chinese Academy of Sciences, Lanzhou 730000, China
\\
{\bf 3} School of Nuclear Science and Technology, University of Chinese Academy of Sciences, Beijing 100049, China
\\
{\bf 4} CAS Key Laboratory of High Precision Nuclear Spectroscopy, Institute of Modern Physics, Chinese Academy of Sciences, Lanzhou 730000, China
\\
{\bf 5} Department of Physics,
Indian Institute of Technology Bombay,Powai, Mumbai 400076,
India\\
*sreeraj@impcas.ac.cn
\end{center}

\begin{center}
\today
\end{center}

% For convenience during refereeing (optional),
% you can turn on line numbers by uncommenting the next line:
%\linenumbers

\definecolor{palegray}{gray}{0.95}
\begin{center}
\colorbox{palegray}{
  \begin{tabular}{rr}
  \begin{minipage}{0.1\textwidth}
    \includegraphics[width=22mm]{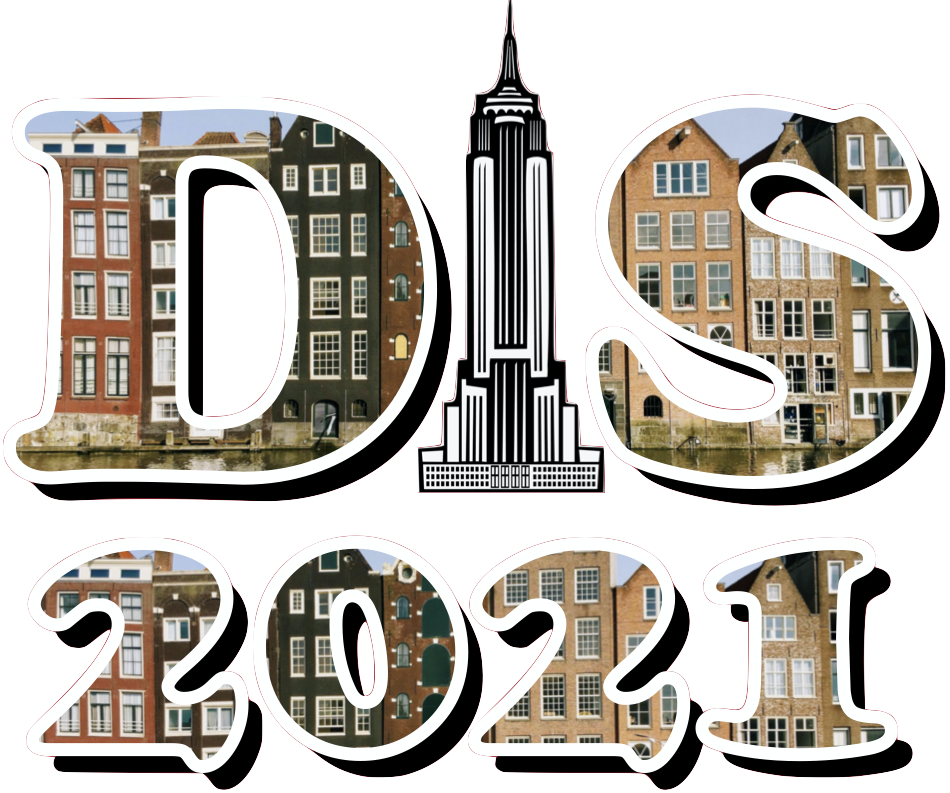}
  \end{minipage}
  &
  \begin{minipage}{0.75\textwidth}
    \begin{center}
    {\it Proceedings for the XXVIII International Workshop\\ on Deep-Inelastic Scattering and
Related Subjects,}\\
    {\it Stony Brook University, New York, USA, 12-16 April 2021} \\
    \doi{10.21468/SciPostPhysProc.?}\\
    \end{center}
  \end{minipage}
\end{tabular}
}
\end{center}

\section*{Abstract}
{\bf
% TODO: write your abstract here.
%The abstract is in boldface, and should fit in 8 lines.
%It should be written in a clear and accessible style, emphasizing the context, the problem(s) studied, the methods used, the results obtained, the conclusions reached, and the outlook. You %can add a table contents, recommended if your paper is more than 6 pages long.

We present a recent calculation of the gravitational form factors (GFFs) of proton using the light-front quark-diquark model constructed by the soft-wall AdS/QCD ~\cite{Chakrabarti:2020kdc}. The four GFFs $~A(Q^2)$ , $B(Q^2)$ , $C(Q^2)$ and $\bar{C}(Q^2)$ are calculated in this model. We also show the  pressure and shear distributions of quarks inside the proton. The GFFs, $A(Q^2)$ and $B(Q^2)$ are found to be consistent with the lattice QCD, while the qualitative behavior of the $D$-term form factor is in agreement with the extracted data from the deeply virtual Compton scattering (DVCS) experiments at JLab, the lattice QCD, and the predictions of different phenomenological models.   
}

%\vspace{10pt}
%\noindent\rule{\textwidth}{1pt}
%\tableofcontents\thispagestyle{fancy}
%\noindent\rule{\textwidth}{1pt}
%\vspace{10pt}

\section{Introduction}
\label{sec:intro}
% TODO: write your article here.
Understanding the parton dynamics inside the nucleon by studying the mechanical properties of the nucleon like the mass, spin and
pressure distribution has recently received widespread attention  following the first experimental evidence of the pressure distribution of quarks inside the nucleon ~\cite{ Burkert:2018bqq}. The matrix elements of the energy-momentum tensor encode the 
information about the gravitational form factors. The gravitational form factors (GFFs) can be assessed experimentally in hard exclusive processes like deeply virtual
Compton scattering (DVCS) through generalized
parton distributions (GPDs)
~\cite{Burkert:2018bqq}.
The symmetric energy-momentum tensor for a spin-half particle is parameterized in terms of four GFFs. The GFFs $A(Q^2)$ and
$B(Q^2)$ are related to the
 mass  and total angular momentum of the proton ~\cite{Ji:1996ek} . 
The GFFs $A$ and $B$ are constrained at $Q^2=0$ as they are related to the generators of the Poincare
group. But the GFF $C(Q^2)$ (also called the D-term) is unconstrained at $Q^2=0$. 
The D-term is connected to the 
internal properties of the nucleon like the pressure and stress
distributions~\cite{Polyakov:2018zvc,Lorce:2018egm}. 

\section{Light-front quark-diquark model}

In the light-front quark-diquark model the three valence quarks inside the proton are reduced to a quark (fermion) and composite state of a diquark (boson). Here we assume that a diquark has spin zero. The light-front wave functions $\psi_{\lambda_q q}^{\lambda_N}(x,\bfk)$
are modeled from the solution of soft-wall AdS/QCD. \cite{Gutsche:2013zia}. The generic ansatz for different spin configurations of nucleon helicities $\lambda_N=\pm \frac{1}{2}$ and for quark helicities $\lambda_q=\pm \frac{1}{2}$ of these light-front wave functions can be written compactly as follows:
\be\label{WF}
\psi_{ \lambda_q q}^{\lambda_N}(x,\bfk)  =   \delta_{\lambda_N,\lambda_q} \varphi_q^{(1)}(x,\bfk)  - \delta_{\lambda_N,-\lambda_q} \varphi_q^{(2)}(x,\bfk) \frac{\bfk}{xM}  e^{i\mathrm{Sign(\lambda_N)}\theta_{\bfk}}
\ee
where, the wave functions $\varphi_q^{(i=1,2)}(x,\bfk)$ are the modified form of the soft-wall AdS/QCD prediction constructed by introducing the parameters $a_q^{(i)}$ and $b_q^{(i)}$ for quark $q$~\cite{BT,Gutsche:2013zia},
\be\label{wf2}
\varphi_q^{(i)}(x,\bfk)&=&N_q^{(i)}\frac{4\pi}{\kappa}\sqrt{\frac{\log(1/x)}{1-x}}x^{a_q^{(i)}}
(1-x)^{b_q^{(i)}}\exp\bigg[-\frac{\bfk^2}{2\kappa^2}\frac{\log(1/x)}{(1-x)^2}\bigg].
\ee

We have taken the quark to be massless and all parameters are set in accordance with the electromagnetic properties of the nucleons. The parameters used in this model are taken from \cite{Mondal:2017wbf}.

\section{Gravitational Form Factors}
The symmetric energy-momentum tensor $T^{\mu \nu}$ for a spin half system can be parameterized in terms of four GFFs  \cite{hari,ji12}
%For a spin $1/2$ composite system, the matrix elements of $T^{\mu \nu}$ involve four gravitational FFs 
\be
\langle P', S'|T^{\mu \nu}_i(0)|P, S\rangle &=&\bar{U}(P', S')\bigg[-B_i(q^2)\frac{\bar{P}^\mu\bar{P^\nu}}{M} +(A_i(q^2)+B_i(q^2))\frac{1}{2}(\gamma^{\mu}\bar{P}^{\nu}+\gamma^{\nu}\bar{P}^{\mu}) \nonumber\\
&+& C_i(q^2)\frac{q^{\mu}q^{\nu}-q^2 g^{\mu\nu}}{M}+\bar{C}_i(q^2)M g^{\mu\nu}\bigg]U(P,S),\label{tensor}
\ee
where, $\bar{P}^\mu=\frac{1}{2}(P'+P)^\mu$, $q^\mu=(P'-P)^\mu$, $U(P,S)$ is the spinor, $Q^2 = -q^2$ and $M$ is the system mass. 
%where $\bar{P}=(P+P')/2$ and $q=P'-P$. 
The GFFs $A_i(Q^2)$ and $B_i(Q^2)$ are calculated using the $T_i^{++}$ component. The GFFs $C_i(Q^2)$ are $\bar{C}_i(Q^2)$ are calculated using the transverse components $T_i^{\perp \perp}$ .
%\be\label{AB}
%\langle P+q, \uparrow|T_i^{++}(0)|P, \uparrow \rangle &=&{2 (P^+)^2} A_i(Q^2), \\
%\langle P+q, \uparrow|T_i^{++}(0)|P, \downarrow \rangle &=& -{2 (P^+)^2}\frac{(q^1_\perp - i q^2_\perp)}{2M}B_i(Q^2)\\
%\langle P+q, \uparrow|T_i^{12}(0)|P, \uparrow \rangle + \langle P+q, \downarrow|T_i^{12}(0)|P, \downarrow \rangle &=& q^1_\perp q^2_\perp \left(A_i(Q^2) + 4 C_i(Q^2) \right) \\ 
%\langle P+q, \uparrow|T_i^{+-}(0)|P, \uparrow \rangle   &=&
%A_i(Q^2) \left( 2 M_n^2 + \frac{q^2}{2}\right) - \frac{q^2}{2} B_i(Q^2) \nn \\ && - q^2 4 C_i(Q^2) + 4 M_n^2\bar{C}_i(Q^2)
%\label{gff1}
%\ee

The distributions of pressure and shear forces inside the nucleon are given by
 \begin{align}
\label{Pre-fun}
p(b)=\frac{1}{6M_n}\frac{1}{b^2}\frac{d}{db} b^2 \frac{d}{db} \tilde{D}(b),\hspace{0.5cm}
s(b)=-\frac{1}{4M_n} b \frac{d}{db} \frac{1}{b} \frac{d}{db} \tilde{D}(b),
\end{align}
where $\tilde{D}(b)$ represents the Fourier transform of GFF $D(Q^2)=4C(Q^2)$ and $b=|{\vec{b}_{\perp}}|$ is the impact parameter.

\begin{figure}[h]
\centering
{\tiny{(a)}}\includegraphics[width=0.47\textwidth]{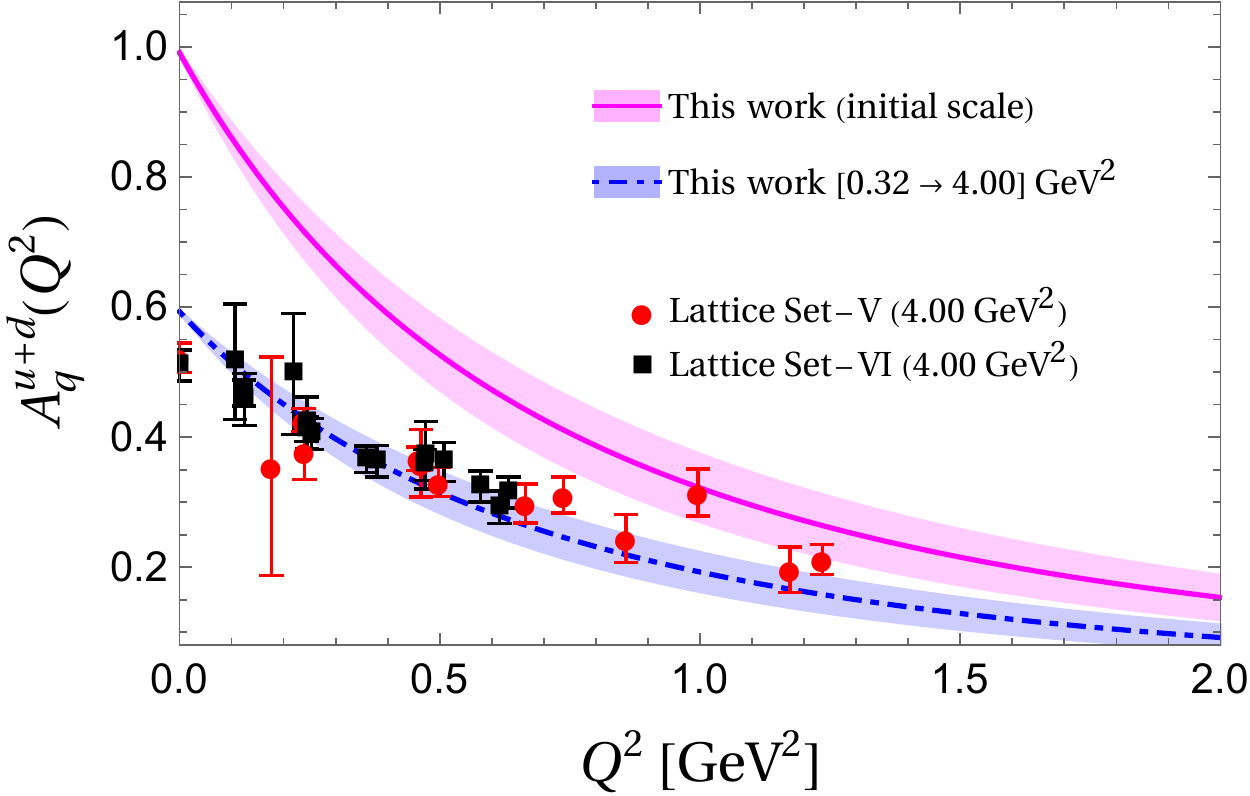}
{\tiny{(b)}}\includegraphics[width=0.47\textwidth]{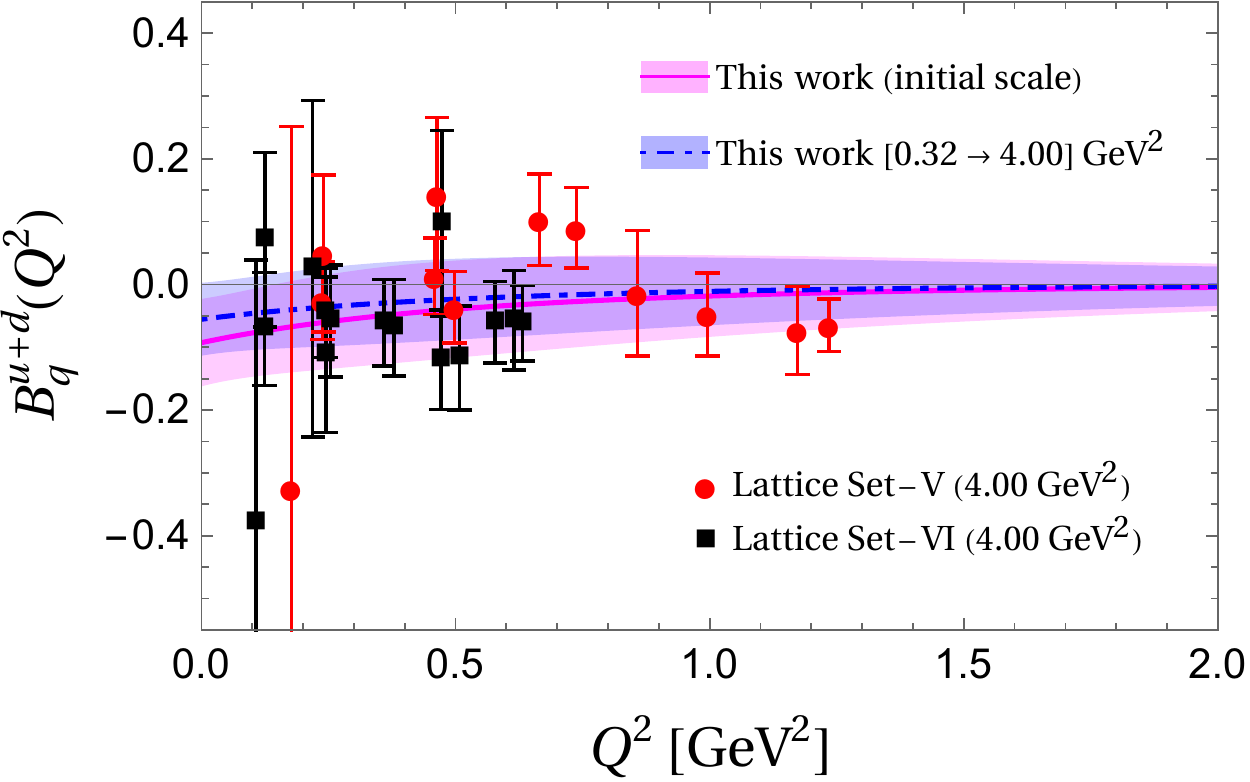}
{\tiny{(c)}}\includegraphics[width=0.47\textwidth]{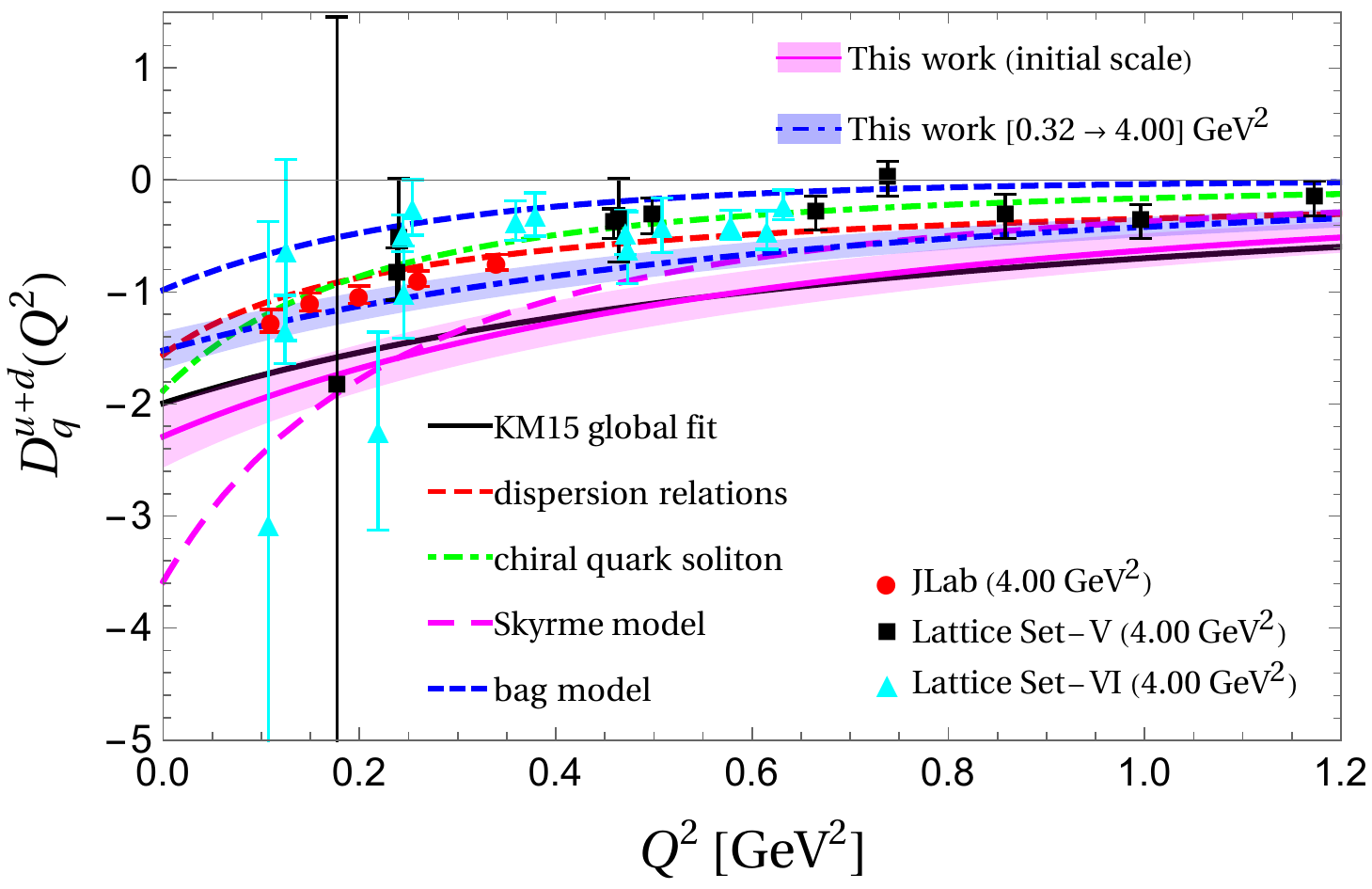}
{\tiny{(d)}}\includegraphics[width=0.47\textwidth]{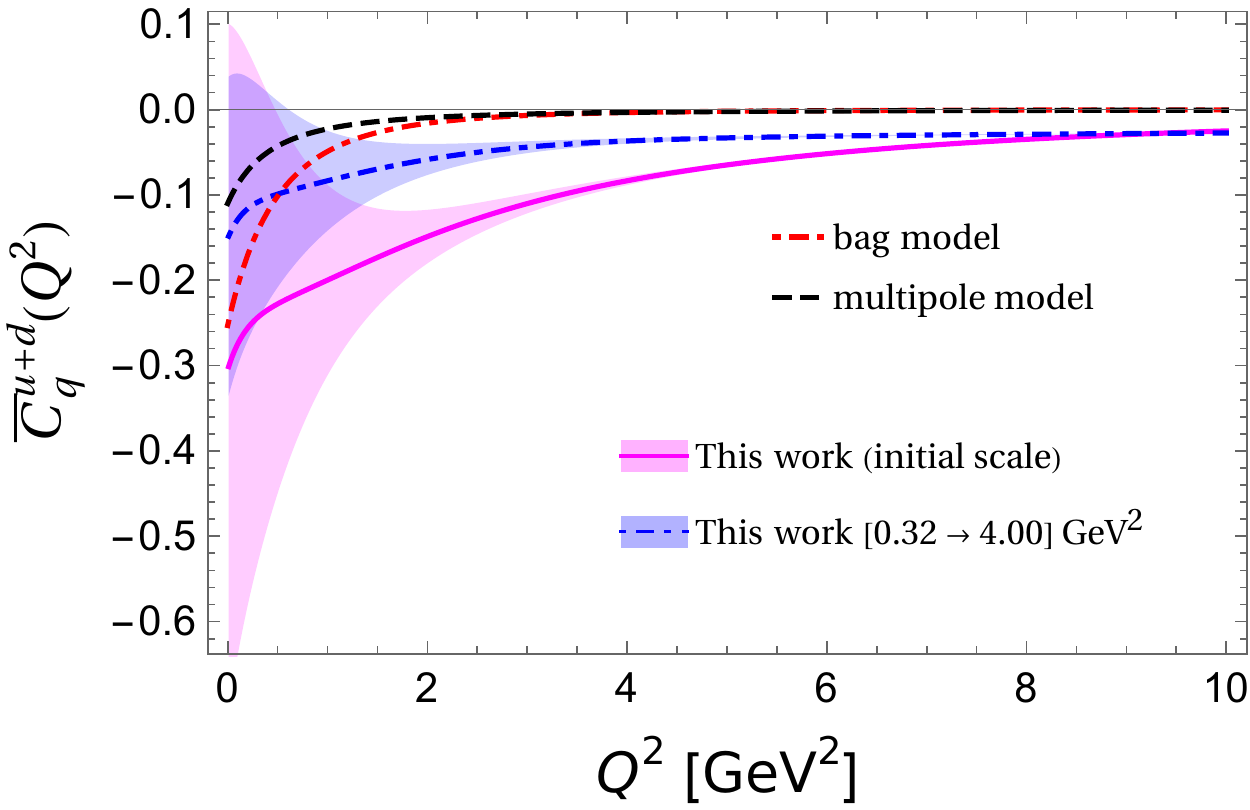}
\caption{The plots of the four GFFs as functions of $Q^2$. The solid magenta lines with magenta bands are the results at the initial scale, while the dash-dotted blue lines with blue bands represent the results at $\mu^2=4$ GeV$^2$ evolved from the initial scale $\mu_0^2=0.32$ GeV$^2$. }
\label{fig1}
\end{figure}

\begin{figure}[h]
\centering
{\tiny{(a)}}\includegraphics[width=0.47\textwidth]{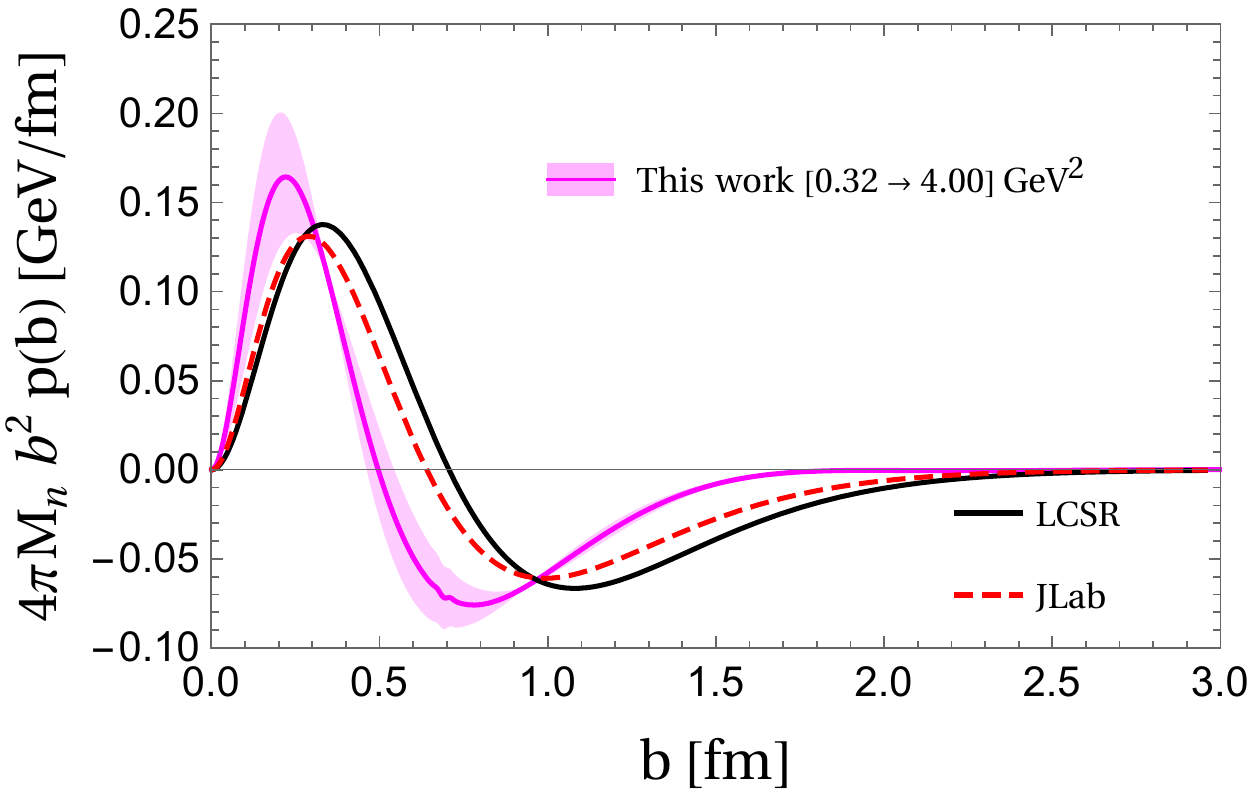}
{\tiny{(b)}}\includegraphics[width=0.47\textwidth]{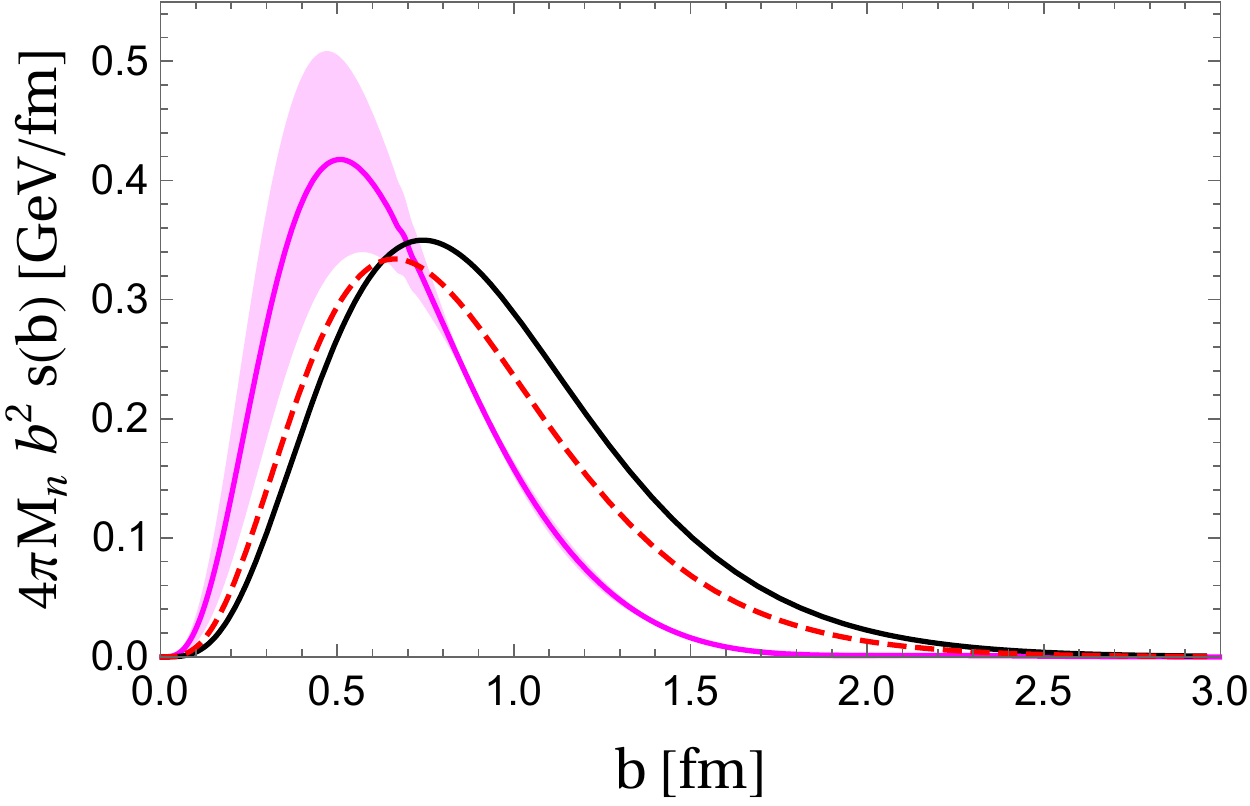}
\caption{Plots of (a) the pressure distribution $4\pi M_nb^2p(b)$, and (b) the shear distribution $4\pi M_nb^2s(b)$ as a function of $b$.}
\label{fig2}
\end{figure}

In Fig.~\ref{fig1} (a) and (b), we show the GFFs $A^{u+d}(Q^2)$ and $B^{u+d}(Q^2)$ which are compared with the lattice QCD prediction~\cite{Hagler:2007xi}. We show the results at the initial scale of $\mu^2_0=0.32$ GeV$^2$  and at the evolved final scale of $\mu^2=4$ GeV$^2$. The evolution is done using the higher order perturbative parton evolution toolkit (HOPPET)~\cite{Salam:2008qg} following Dokshitzer-Gribov-Lipatov-Altarelli-Parisi (DGLAP) equations of QCD with next-to-next-to-leading order. We observe that after QCD evolution, $A^{u+d}(Q^2)$ and $B^{u+d}(Q^2)$ are  consistent with the lattice QCD results. In Fig.~\ref{fig1} (c) we show our result for the GFF $4C^{u+d}(Q^2) = D^{u+d}(Q^2)$.  The evolved value at zero momentum transfer is $D^{u+d}(0) = -1.521$. The fitted function $\frac{-1.521}{(1 + 0.531 \mathrm{Q}^2)^{3.026}}$ reproduces the evolved result for $D^{u+d}(Q^2)$. 
We find that the qualitative behavior of our result is compatible with lattice QCD~\cite{Hagler:2007xi} and the experimental data from JLab~\cite{Burkert:2018bqq} as well as other theoretical predictions from the KM15 global fit~\cite{Kumericki:2015lhb}, dispersion relation~\cite{Pasquini:2014vua}, $\chi$QSM~\cite{Goeke:2007fp}, Skyrme model~\cite{Cebulla:2007ei}, and bag model~\cite{Ji:1997gm}. In Fig.~\ref{fig1} (d) our result for the GFF  $\bar{C}^{u+d}(Q^2)$ is shown. The error bands reflect a $10\%$ uncertainty in the model parameters. In Fig.~\ref{fig2} (a) and (b) we show the profile for the pressure and shear distributions of the quark inside the nucleon in the impact parameter space. Our results are compared with the distribution evaluated in leading order light-cone sum rule~\cite{Anikin:2019kwi} and the distribution obtained from the fitting functions of the experimental data for $D(Q^2)$ at JLab~\cite{Burkert:2018bqq}. The pressure distribution satisfies the von Laue stability condition ~\cite{VonL} showing a positive core and a negative tail. The shear distribution for stable hydrostatic systems are positive and has connection to surface tension and surface energy ~\cite{Polyakov:2018zvc}. We observe that our shear distribution is also positive and the qualitative behavior is in accord with other approaches~\cite{Anikin:2019kwi,Goeke:2007fp,Cebulla:2007ei}.

\section{Conclusion}
We calculated the four GFFs of a spin half nucleon in the light-front scalar quark-diquark model. The D-term was extracted using the perpendicular component of the energy momentum tensor. Our results for $A(Q^2)$ and $B(Q^2)$ were found to be consistent with lattice QCD predictions. The evolved result for the D-term is in accord with the experimental data extracted from DVCS process at JLab. The value of $\bar{C}(Q^2)$ is found to be negative at zero momentum transfer. Our pressure $p(b)$ and shear force $s(b)$ distributions are found to be consistent with the experimental observation and other theoretical predictions.

\section*{Acknowledgements}
SN thanks the organizers of ``XXVIII International Workshop on Deep-Inelastic Scattering and Related Subjects'' for the
opportunity to present this work. C. M. and S. N. thank the Chinese Academy of Sciences Presidents International Fellowship Initiative for the support via Grants No. 2021PM0023 and 2021PM0021, respectively. X. Z. is supported by new faculty startup funding by the Institute of Modern Physics, Chinese Academy of Sciences, by Key Research Program of Frontier Sciences, Chinese Academy of Sciences, Grant No. ZDB-SLY-7020, by the Natural Science Foundation of Gansu Province, China, Grant No. 20JR10RA067 and by the Strategic Priority Research Program of the Chinese Academy of Sciences, Grant No. XDB34000000.

% \bibliographystyle{SciPost_bibstyle} % Include this style file here only if you are not using our template
\bibliography{SciPost_Example_BiBTeX_File.bib}

\nolinenumbers

\end{document}